\documentclass[amsmath,amssymb,prd,twocolumn]{revtex4-2}
\usepackage{dcolumn}
\pdfoutput=1
\usepackage[english]{babel}
\usepackage[utf8]{inputenc}
\usepackage[T1]{fontenc}
\usepackage{amsthm}
\usepackage{amsfonts}
\usepackage{url}
\usepackage{tensor}
\usepackage[colorlinks = true,
            linkcolor = blue, 
            linkcolor = blue,
            urlcolor  = blue,
            citecolor = blue,
            anchorcolor = blue]{hyperref}
\usepackage{graphicx}
\usepackage{mathrsfs}
\usepackage{enumitem}
\usepackage{caption}
\usepackage{autonum}

\usepackage[usenames,dvipsnames]{xcolor}

\newcounter{mnotecount}[section]

\newcommand{\beq}{\begin{eqnarray}}
\newcommand{\eeq}{\end{eqnarray}}
\newcommand{\ben}{\begin{eqnarray*}}
\newcommand{\een}{\end{eqnarray*}}

\newtheorem*{theorem*}{Theorem}
\theoremstyle{definition}

\newcommand\ringring[1]{%
  {
   \mathop{\kern0pt #1}\limits^{
     \vbox to-1.85ex{
       \kern-2ex 
       \hbox to 0pt{\hss\normalfont\kern.1em \r{}\kern-.45em \r{}\hss}
       \vss 
     }   }
  }
}

\begin{document}
\title{Halilsoy and Chandrasekhar standing gravitational waves in the linear approximation}
\author{Kornelia Nikiel}
\author{Sebastian J. Szybka}
\affiliation{Astronomical Observatory, Jagiellonian University}
\begin{abstract}
Halilsoy and Chandrasekhar cylindrical standing gravitational waves correspond to two different classes of solutions to the vacuum Einstein equations.  Both families satisfy the definition of standing gravitational waves proposed by Stephani, but only the latter class fulfills the stricter definition introduced by Chandrasekhar. The aim of this research is to compare both classes of solutions within the linear regime. We discover that the linearized Halilsoy and Chandrasekhar standing waves are gravitational analogs of two different types of electromagnetic polarization standing waves.

\end{abstract}
\maketitle{}

\section{Introduction}

The standing waves play important role in many fields of physics. In general relativity, the nonlinearity of the Einstein equations makes gravitational standing waves an interesting but difficult subject of study.  In a cylindrical symmetry, the Einstein equations simplify considerably and the problem of standing gravitational waves may be approached in its full nonlinear form. 

A standing wave is a wave that does not transport energy in space. In general relativity, a density of gravitational energy cannot be defined locally which makes a definition of standing gravitational waves problematic. In cylindrical symmetry, the problem was addressed independently by Hans Stephani \cite{Stephani:2003} and Subrahmanyan Chandrasekhar \cite{chandra} using the concept of C-energy \cite{thorne}. A definition proposed by Stephani \cite{Stephani:2003} is less strict than the assumption adopted by Chandrasekhar \cite{chandra}. In Chandrasekhar approach, C-energy remains constant in time. According to Stephani, C-energy should be constant in time only on average. 

In this paper, we compare these two classes of standing gravitational waves within the linear regime. This approach allows us to clarify their differences and to find their weak field interpretation as gravitational analogs of electromagnetic polarization standing waves. Furthermore, the linearization sheds light on the nuances concerning the definitions of nodes and antinodes.

\section{Halilsoy waves}

In geometrized units the cylindrically symmetric metric can be written in the form

\begin{equation}\label{metricH}
g=e^{2(\gamma-\psi)}\left(-dt^2+d\rho^2\right)+\rho^2e^{-2\psi}d\phi^2+e^{2\psi}(dz+\omega d\phi)^2\;,
\end{equation}
where $\rho>0$, $-\infty<t,z<\infty$, and $0\leq\phi<2\pi$. The coordinates are adapted to the Killing fields $\partial_{z}$, $\partial_\phi$, hence the metric functions $\psi$, $\gamma$, and $\omega$ depend on $t$ and $\rho$ only.  If $\omega=0$, the metric \eqref{metricH} with appropriate choice of the \mbox{metric} functions reduces for vacuum to the Einstein-Rosen waves \cite{EinsteinRosen37,Rosen54}.

Halilsoy showed \cite{Halilsoy:1988} that the simplest Einstein-Rosen standing wave solution  may be extended to include a second polarization mode. The exact solution to the Einstein equations has a form
\begin{equation}\label{solH}
\begin{split}
	e^{-2\psi}=&e^{A J_0 \cos{(t/\lambda)}}\sinh^2\frac{\alpha}{2}+e^{-AJ_0 \cos{(t/\lambda)}}\cosh^2\frac{\alpha}{2}\;,\\
	\omega=&-(A\sinh\alpha)\,\rho\, J_1\sin{( t/\lambda)}\;,\\
	\gamma=&\frac{1}{8}A^2\left[\left(\frac{\rho}{\lambda}\right)^2(J_0^2+J_1^2)-2\frac{\rho}{\lambda} J_0 J_1\cos^2(t/\lambda)\right]\;,
\end{split}
\end{equation}
where $\lambda> 0$, $A$, $\alpha$ are constants, and \mbox{$J_0=J_0(\rho/\lambda)$} and \mbox{$J_1=J_1(\rho/\lambda)$} are the Bessel functions of the first kind and orders $0$ and $1$, respectively.

The metric \eqref{metricH} with the auxiliary functions defined above constitutes the regular exact solution to the vacuum Einstein equations. For $\alpha=0$ it reduces to the simplest Einstein-Rosen standing wave solution which has been recently investigated in the article \cite{szybkanaqvi}.

\section{Chandrasekhar waves}

The alternative form in which cylindrical gravitational waves can be described is given by the metric
\begin{equation}\label{metricCh}
	\hat{g}=e^{2\nu}\left(-dt^2+d\rho^2\right)+\frac{\rho^2}{\Psi}d\phi^2+\Psi(dz+\Omega d\phi)^2\;,
\end{equation}
where the coordinates have a standard range and where $\Omega$ corresponds to $-q_2$ in Chandrasekhar notation \cite{chandra}.
The standing wave solution investigated in the Chandrasekhars' paper is given by
\begin{equation}
\Psi=\frac{1-F^2}{1+F^2-2F\cos(t/\lambda)}\;,
\end{equation}
and
\begin{equation}
	\Omega/\lambda=\frac{2\rho F_{,\rho}}{1-F^2}\cos{(t/\lambda)}+\frac{4}{\lambda^2}\int_0^\rho\frac{\hat\rho F^2}{(1-F^2)^2}d\hat\rho\;,
\end{equation}
where $F=F(\rho)$. 

In cylindrical symmetry, mass cannot be defined in the standard way. To address this issue, Kip Thorne introduced the concept of C-energy \cite{thorne}. Further insights into C-energy are provided by Chandrasekhar \cite{chandra}. For the solution under investigation C-energy equals to $\nu+\ln\sqrt\Psi$. Chandrasekhar assumed that the time-independence of C-energy is a necessary condition for the existence of standing waves. The Halilsoy solution \eqref{metricH} satisfies this condition only after averaging in time (C-energy is equal to $\gamma$). For the Halilsoy solution C-energy is strictly constant only at zeros of the Bessel functions $J_0(\rho/\lambda)$ and $J_1(\rho/\lambda)$.

In the case studied by Chandrasekhar, C-energy is time-independent and the Einstein equations simplify to
\begin{equation}
\begin{split}
	(\nu+\ln\sqrt\Psi)_{,t}=&0\;,\\
	(\nu+\ln\sqrt\Psi)_{,\rho}=&\frac{\rho}{(1-F^2)^2}(F^2/\lambda^2+(F_{,\rho})^2)\;,
\end{split}
\end{equation}
where the essential part of the Einstein equations has the form
\begin{equation}\label{F}
F(1+F^2)/\lambda^2+\frac{1-F^2}{\rho}(\rho F_{,\rho})_{,\rho}+2F(F_{,\rho})^2=0\;,
\end{equation}
with the boundary conditions $F(0)=F_0$, $F_{,\rho}(0)=0$, where $0<F_0<1$ is a constant. The constant \mbox{$\lambda>0$} is a parameter of the solution. No exact nontrivial solutions to the equation above are known so it must be solved numerically. The function $\Psi$ can be determined directly from $F$. The remaining metric functions $\nu$ and $\Omega$ are calculated via quadratures. The Chandrasekhar solutions are regular.

\section{Linear approximation}

\subsection{Halilsoy waves}

The inspection of the line element \eqref{metricH} reveals that $\psi=\gamma=0$ (implied by $A=0$) correspond to Minkowski spacetime in cylindrical coordinates provided that a trivial transformation of coordinates is applied: $t\rightarrow t/\sigma$, $\rho\rightarrow \rho/\sigma$, $z\rightarrow \sigma z$, $\lambda\rightarrow \lambda/\sigma$, where $\sigma$ is a dimensionless constant given by $$\sigma=\sqrt{\sinh^2\frac{\alpha}{2}+\cosh^2\frac{\alpha}{2}}=\sqrt{\cosh{\alpha}}\;.$$ The linearization of the metric \eqref{metricH} with $\alpha=0$ (Einstein-Rosen waves) around Minkowski spacetime in an appropriate gauge is a textbook exercise (see the exercise $35.3$ in the book \cite{MTW}). 

After some algebra, the nonzero perturbations of the metric \eqref{metricH} (linear in $\epsilon=A/\sigma^2$) can be written as (nonzero components)
\begin{equation}
\begin{split}
	h_{tt}=h_{zz}=-h_{xx}=-h_{yy}=&\epsilon Q_0\;,\\
	h_{xz}=h_{zx}=&\epsilon y \,\sinh(\alpha)W/\rho\;,\\
	h_{yz}=h_{zy}=&-\epsilon x \,\sinh(\alpha)W/\rho\;,\\
\end{split}
\end{equation}
where 
\begin{equation}\label{QW}
	\begin{split}
	Q_i=&J_i(\rho/\lambda)\cos(t/\lambda)\;,\\
		W=& J_1(\rho/\lambda)\sin(t/\lambda)=-\omega/(\sinh(\alpha)\,\epsilon\,\sigma\rho)\;,
\end{split}
\end{equation}
and $\rho=\sqrt{x^2+y^2}$. For $\rho\longrightarrow 0$ nondiagonal components of $h_{\mu\nu}$ vanish.

In order to compare linearized Halilsoy and Chandrasekhar solutions, we need to use the same gauge. Since both solutions describe waves, they should have the simplest form in the TT gauge. Only pure waves can be reduced into the TT gauge. The linearized Halilsoy solution is a pure wave, namely, it is easy to verify that $\Box h_{\mu\nu}=0$.

We find $h=-2\epsilon Q_0$, hence the trace-reverse tensor $\bar{h}_{\mu\nu}=h_{\mu\nu}-\frac h 2 \eta_{\mu\nu}$ has the form
\begin{equation}\label{h1}
\begin{split}
	\bar{h}_{tt}=\bar{h}_{xx}=\bar{h}_{yy}=&0\;,\\
	\bar{h}_{zz}=&2\epsilon Q_0\;,\\
	\bar{h}_{xz}=\bar{h}_{zx}=&\epsilon y \,\sinh(\alpha)W/\rho\;,\\
	\bar{h}_{yz}=\bar{h}_{zy}=&-\epsilon x \,\sinh(\alpha)W/\rho\;.\\
\end{split}
\end{equation}

The Riemann tensor is invariant under gauge transformations. Since in the TT gauge $R_{i0k0}=-\frac{1}{2} h^{TT}_{ik,00}$, then it is sufficient to calculate the components of the Riemann tensor in our gauge and integrate them twice to find $h^{TT}_{ik}$. Using the standard relation $$R_{\alpha\beta\mu\nu}=\frac{1}{2}(h_{\alpha\nu,\beta\mu}+h_{\beta\mu,\alpha\nu}-h_{\alpha\mu,\beta\nu}-h_{\beta\nu,\alpha\mu})$$ and identity $Q_2=2\lambda/\rho\, Q_1-Q_0$ we obtain
\begin{equation}\label{h2}
\begin{split}
	h^{TT}_{xx} =& -\frac{\epsilon}{\rho^2}\left[y^2\, Q_0 +\lambda (x - y)(x + y) Q_1/\rho \right]  \;, \\
	h^{TT}_{yy} =& -\frac{\epsilon}{\rho^2}\left[x^2\, Q_0 -\lambda (x - y)(x + y) Q_1/\rho \right]  \;, \\
	h^{TT}_{zz} =& \epsilon \, Q_0\;, \\
	h^{TT}_{xy} = h^{TT}_{yx}=& \frac{\epsilon }{\rho^2} x y\left[Q_0 - 2\lambda Q_1/\rho\right]\;, \\
	h^{TT}_{xz} = h^{TT}_{zx}=& \epsilon y\,\sinh(\alpha) W/\rho\;,\\
	h^{TT}_{yz} = h^{TT}_{zy}=& -\epsilon x\, \sinh(\alpha) W/\rho\;.
\end{split}
\end{equation}
It is instructive to find these components along $x$-axis ($y=0$)  
\begin{equation}\label{h3}
	h^{TT}_{\alpha\beta}|_{y=0} = \left(
\begin{array}{cccc}
0 & 0 & 0 & 0 \\[2ex]
	0 & -\epsilon \lambda  Q_1/x & 0  & 0 \\[2ex]
	0 & 0 & -\epsilon \left[Q_0-\lambda Q_1/x \right]& -\epsilon\sinh(\alpha)W \\[2ex]
	0 & 0 &   -\epsilon\sinh(\alpha)W  & \epsilon \, Q_0 \\
\end{array}
\right)\;.
\end{equation}

Finally, we notice that all components of the perturbation tensor are well-behaved at the axis $\rho=0$ because $\lim_{\rho\rightarrow 0^+}J_1(\rho/\lambda)/\rho=\frac{1}{2\lambda}$.


\subsection{Chandrasekhar waves}

We repeat the same procedure for the Chandrasekhar solution.
The line element \eqref{metricCh} with $F=0$ corresponds to Minkowski spacetime in cylindrical coordinates ($\Psi=1$, $\Omega=0$, and $\nu=const$, where without loss of generality we may assume $\nu=0$). In general, the essential part of the Einstein equations, namely the Eq.\ \eqref{F}, can be solved only numerically. The linearized version takes the form of a Bessel equation of order zero
\begin{equation}
	F/\lambda^2+\frac{F_{,\rho}}{\rho}+F_{,\rho\rho}=0\;.
\end{equation}
The solution to this equation is given by a linear combination of the Bessel functions of the first kind and the second kind and zero order, namely $J_0$ and $Y_0$
\begin{equation}
	F(\rho)=c_1 J_0(\rho/\lambda)+c_2 Y_0(\rho/\lambda)\;.
\end{equation}
Following Chandrasekhar \cite{chandra}, we observe that the regularity at the axis $\rho=0$ implies $F'(0)=0$, hence we choose $c_2=0$. Since $F$ should be small, we set $c_1=\epsilon/2$. The linearized Chandrasekhar waves can be derived from $F(\rho)=\frac{\epsilon}{2} J_0(\rho/\lambda)+O(\epsilon^2)$. The nonzero components of the perturbation $h'_{\alpha\beta}$ of the metric \eqref{metricCh} are

\begin{equation}
\begin{split}
	h'_{tt}=h'_{zz}=-h'_{xx}=-h_{yy}=&\epsilon Q_0\;,\\
	h'_{xz}=h'_{zx}=&\epsilon y \,Q_1/\rho\;,\\
	h'_{yz}=h'_{zy}=&-\epsilon x \,Q_1/\rho\;,\\
\end{split}
\end{equation}
where auxiliary functions $Q_i$ are given by the Eq.\ \eqref{QW}.

We first confirm that $\Box h'_{\mu\nu}=0$ and compute the trace-reversed tensor, then proceed to the TT gauge.  We find $h'=-2\epsilon Q_0$. The trace-reverse tensor $\bar{h}'_{\mu\nu}$ has the form (nonzero components)
\begin{equation}\label{hp1}
\begin{split}
        \bar{h}'_{tt}=\bar{h}'_{xx}=\bar{h}'_{yy}=&0\;,\\
        \bar{h}'_{zz}=&2\epsilon Q_0\;,\\
        \bar{h}'_{xz}=\bar{h}'_{zx}=&\epsilon y\, Q_1/\rho\;,\\
        \bar{h}'_{yz}=\bar{h}'_{zy}=&-\epsilon x\, Q_1/\rho\;.
\end{split}
\end{equation}
With the help of the Riemann tensor, we find the components of the TT gauge metric perturbation
\begin{equation}\label{hp2}
\begin{split}
        h'^{TT}_{xx} =& -\frac{\epsilon}{\rho^2}\left[y^2\, Q_0 +\lambda (x - y)(x + y) Q_1/\rho \right]  \;, \\
        h'^{TT}_{yy} =& -\frac{\epsilon}{\rho^2}\left[x^2\, Q_0 -\lambda (x - y)(x + y) Q_1/\rho \right]  \;, \\
        h'^{TT}_{zz} =& \epsilon \, Q'_0\;, \\
        h'^{TT}_{xy} = h'^{TT}_{yx}=& \frac{\epsilon }{\rho^2} x y\left[Q_0 - 2\lambda Q_1/\rho\right]\;, \\
        h'^{TT}_{xz} = h'^{TT}_{zx}=& \epsilon  y\, Q_1/\rho\;,\\
        h'^{TT}_{yz} = h'^{TT}_{zy}=& -\epsilon  x\, Q_1/\rho\;.
\end{split}
\end{equation}
We notice that $h'^{TT}_{\alpha\beta}$ are well-behaved at the axis $\rho=0$.
Again, it is instructive to find these components along $x$-axis ($y=0$)
\begin{equation}\label{hp3}
	h'^{TT}_{\alpha\beta}|_{y=0} = \left(
\begin{array}{cccc}
0 & 0 & 0 & 0 \\[2ex]
        0 & -\epsilon \lambda  Q_1/x & 0  & 0 \\[2ex]
        0 & 0 & -\epsilon \left[Q_0-\lambda Q_1/x \right]& -\epsilon  \,Q_1 \\[2ex]
0 & 0 &   -\epsilon \,Q_1 & \epsilon \, Q_0 \\
\end{array}
\right)\;.
\end{equation}
The formula above can be easily compared to the formula \eqref{h3}.

\subsection{Comparison}

The linearized Halilsoy and Chandrasekhar solutions have similar forms in the TT gauge. The essential difference is hidden in the components of the metric related to polarization. The formulas for the Chandrasekhar solution \eqref{hp1}, \eqref{hp2}, \eqref{hp3} differ from those for the Halilsoy solution \eqref{h1}, \eqref{h2}, \eqref{h3} solely in their nondiagonal components. The expression $\sinh(\alpha)W$ that appears in the Halilsoy solution is replaced by $Q_1$ in the Chandrasekhar solution. The Halilsoy solution for $\alpha=0$ is linearly polarized (the metric function $\omega$ vanishes). The Chandrasekhar solution has always nontrivial polarization term $\Omega$ whose time-dependence is shifted by $\pi/2$ in a phase relative to time-dependence of the term $\omega$ for the Halilsoy solution. The inspection of formulas \eqref{h3}, \eqref{hp3} reveals that far away from the center (the large $x$) both waves become superpositions of plane waves with a supressing factor in amplitudes.

\subsubsection{Test sphere}

We compare below how both gravitational waves act on a small sphere of test particles. Without loss of generality we assume that the sphere is centered at $y=z=0$ plane. By a small sphere we mean a collection of particles  with coordinates $(x,y,z)$ such that $(x-x_0)^2+y^2+z^2=r^2$ at the spatial $t=const$ hypersurfaces. 

Let $\hat n=(\cos\phi\sin\theta,\sin\phi\sin\theta,\cos\theta)$ be a vector pointing from the origin $(x_0,0,0)$ to a point on the sphere (a unit vector in the unperturbed metric). The spatial distance from the center to this point is given by a length of a radial curve 
$$\gamma^i(l)=(x_0+l\sin\theta\cos\phi,l\sin\theta\sin\phi,l\cos\theta)\;,$$
which is given by the formula
$$\Delta s(\hat n)=\int_\gamma ds=\int_0^r \sqrt{g_{ij}n^i n^j}dl\;,$$ where the angles remain constant along the curve. We have $g_{ij}=\delta_{ij}+h^{TT}_{ij}$. 
For both solutions we find
\begin{equation}\label{deltaS}
\begin{split}
	\Delta s(\hat n)=&r\left[1+\frac{\epsilon}{2}\big(Q_0(\cos^2\theta-\sin^2\theta \sin^2\phi)\right.\\
	&\left.-K\sin2\theta\sin\phi-\frac{\lambda}{x_0}Q_1\sin^2\theta\cos2\phi\big)\right]\;,
\end{split}
\end{equation}
where $K=\sinh(\alpha)W$ for the Halilsoy solution and \mbox{$K=Q_1$} for the Chandrasekhar solution. The precise value of a small constant $r$ is not important in our considerations. In this formula, $x$ (or $x_0$)  can always be replaced by $\rho$ through an appropriate rotation of the coordinate system, which we will silently assume whenever convenient in the remainder of the text. 

In some special cases, the formula \eqref{deltaS} simplifies considerably because the spatial amplitude separates from the time-dependent factor 
\begin{equation}\label{fac}
\Delta s(\hat n)=r[1+\epsilon P(t/\lambda)H(x_0,\theta,\phi)]\;,
\end{equation}
where $P(t)=\cos(t/\lambda)$ or $P(t)=\sin(t/\lambda)$ and where $H(x_0,\theta,\phi)$ denotes the spatial amplitude of the distortion. Such separation holds for the Chandrasekhar solution. It is also valid for the Halilsoy solution in the following cases
\begin{enumerate}[label=(\roman*)]
\item for $\alpha=0$ (the Einstein-Rosen standing wave),
\item in an approximate form for a large $\alpha$,
\item at the center $x_0=0$,
\item if $\sin2\theta=0$ or $\sin\phi=0$.
\end{enumerate}

\begin{figure}[t!]
	\includegraphics[width=0.745\linewidth,angle=0]{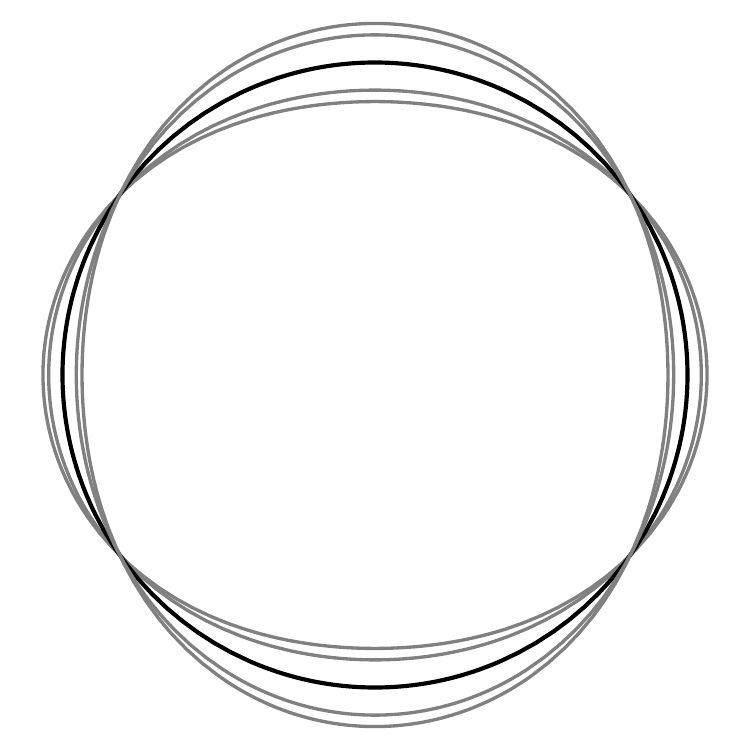}
	\caption{The deformation of a cross section $y=0$ of a test sphere by the Halilsoy and Chandrasekhar standing gravitational waves. The sphere is centered at the origin and plotted for $t/\lambda\in\{0,\pi/4,\pi/2,3\pi/4,\pi\}$. The thick ring represents the unperturbed sphere at $t/\lambda=\pi/2$. The figure is invariant under rotations in $\phi$ (the horizontal direction). The deformation is exaggerated for clarity ($\epsilon=1/2$).}
\label{fig1}
\end{figure}

If the sphere is centered at the origin, then the formula does not depend on $\alpha$ and for both solutions simplifies to
\begin{equation}
	\Delta s(\hat n)_{x_0=0}=r\left\{1+\frac{\epsilon}{8}\left[1+3\cos(2\theta)\right]\cos(t/\lambda)\right\}\;,
\end{equation}
where we used the fact that $\lim_{x_0\rightarrow 0^+}J_1(x_0)/x_0=1/2$. The deformation of a cross section of the centered test sphere is presented in Fig.\ \ref{fig1}.

\subsubsection{Polarization}

We analyze the deformation of a test sphere far from the center. 

\begin{figure}[t!]
	\includegraphics[width=0.745\linewidth,angle=0]{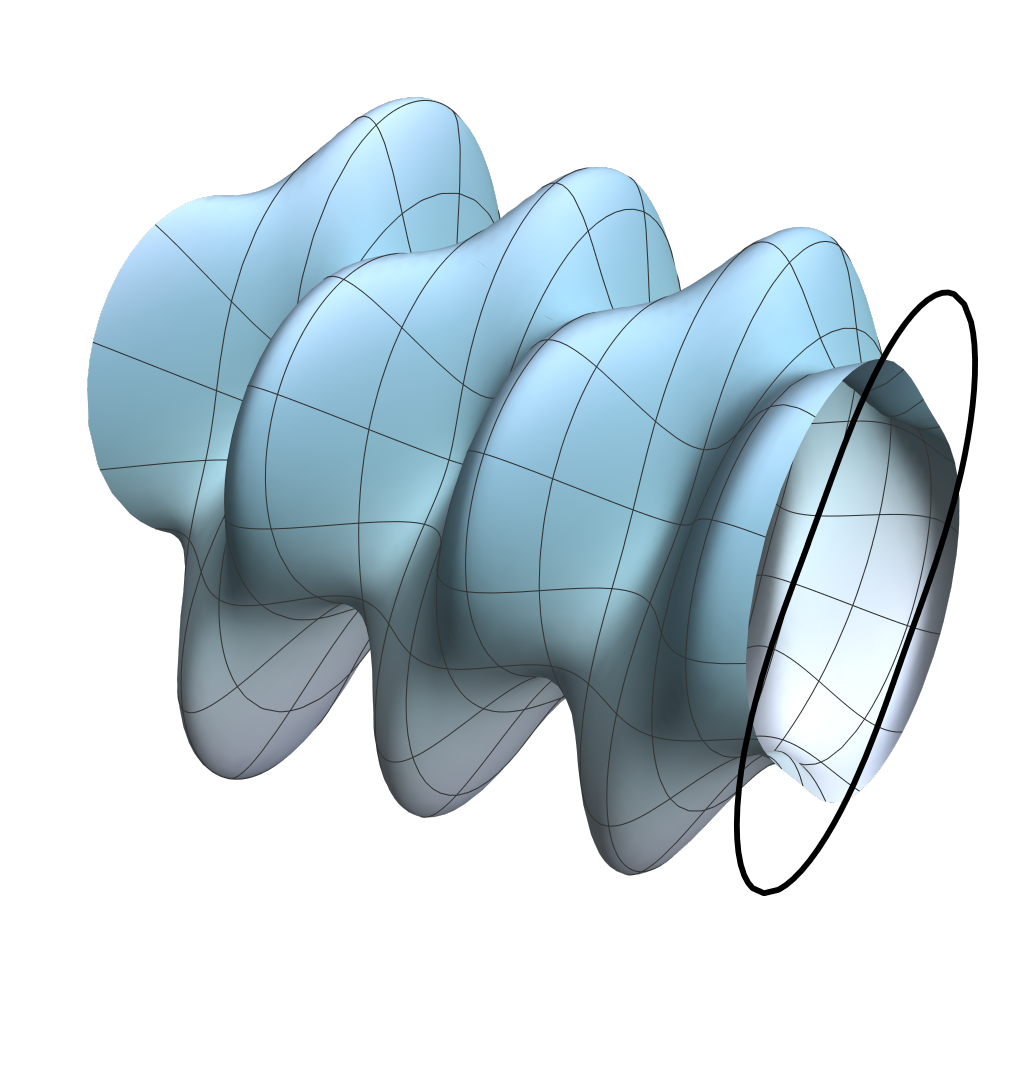}
	\caption{The deformation of a test tube along $\partial_\rho$ direction by the Halilsoy standing gravitational wave. The snapshot is taken at $t/\lambda=7\pi/4$ for $\alpha=2$ and it extends from $\rho/\lambda=2\pi$ to $\rho/\lambda=8\pi$. The black deformed ring at $\rho/\lambda=8\pi$ corresponds to the cross section of the tube at $t/\lambda=11\pi/4$. The deformation is exaggerated for clarity ($\epsilon=3/2$).}
\label{fig2}
\end{figure}

\begin{figure}[t!]
	\includegraphics[width=0.745\linewidth,angle=0]{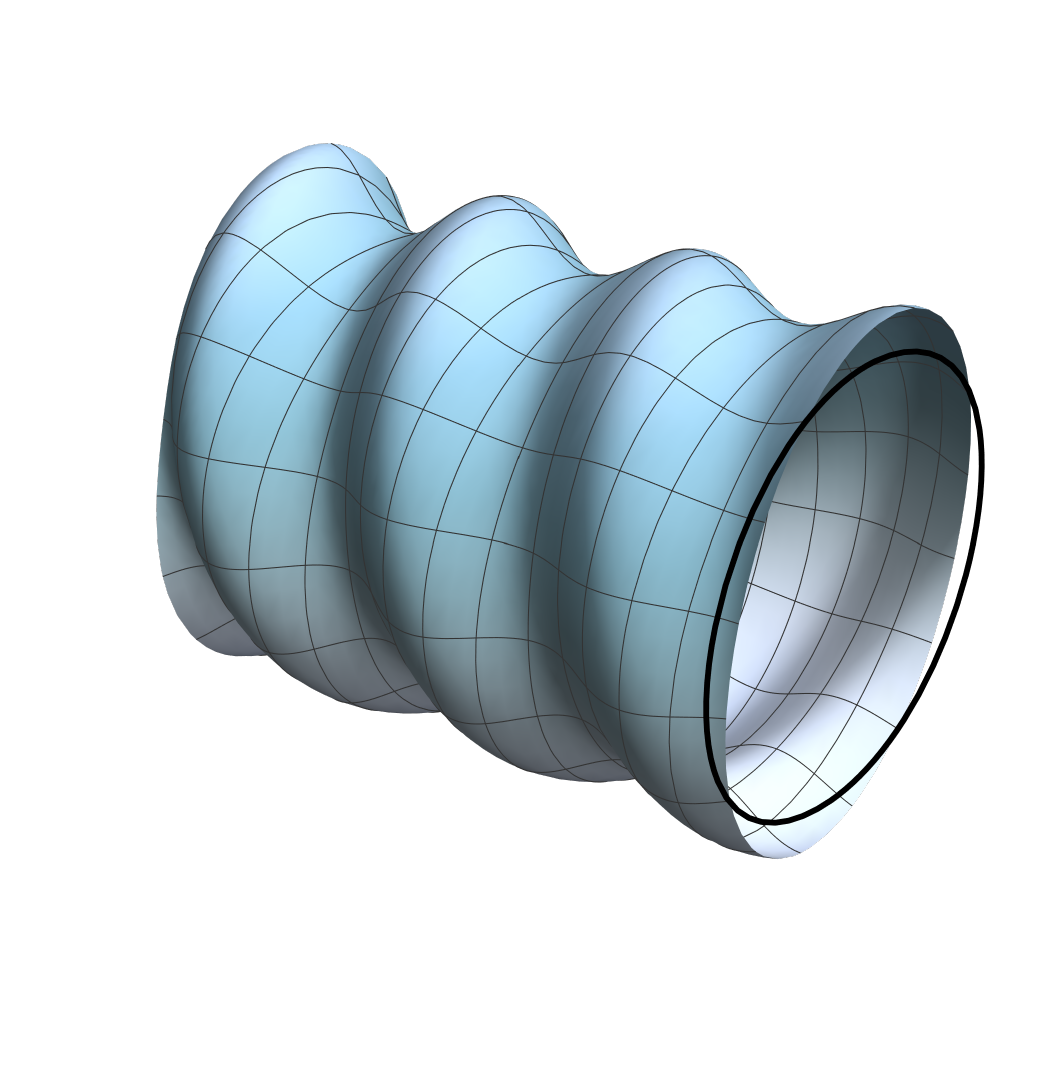}
	\caption{The deformation of a test tube along $\partial_\rho$ direction by the Chandrasekhar standing gravitational wave. The snapshot is taken at $t/\lambda=7\pi/4$ and it extends from $\rho/\lambda=2\pi$ to $\rho/\lambda=8\pi$. The black deformed ring at $\rho/\lambda=8\pi$ corresponds to the cross section of the tube at $t/\lambda=11\pi/4$. The deformation is exaggerated for clarity ($\epsilon=3/2$). The right-handedness of the polarization along $\partial_\rho$ is clearly visible.}
\label{fig3}
\end{figure}

A cross section of the test sphere along a plane orthogonal to the direction of propagation of waves ($\phi=\pi/2$) seen as a function of $\rho$ forms a tube. Figures \ref{fig2}, \ref{fig3} show how this tube may be deformed by the standing waves (see enclosed animations \cite{AnimLSGW}). For the Halilsoy waves the form of deformations depends strongly on the parameter $\alpha$ as will be explained later in the text.

The four directions of the extremal distortions of the tube in the plane orthogonal to $\partial_\rho$ can be determined from the equation $\frac{\partial\Delta s(\hat n)|_{\phi=\pi/2}}{\partial \theta}=0$. The exact solutions to this equation can be found, but the formulas are too long to be usefully cited here. We denote solutions to this equation by $\theta_{ex}$ and analyze special cases below.

For the Halisoy solution for $\alpha=0$ the spatial amplitude separates from a time-dependent factor \eqref{fac} and the distortion $\Delta s-r$ is proportional to $\cos(t/\lambda)$. At $t/\lambda=\pi/2+ k\pi$, where $k\in\mathbb{Z}$ the tube is not deformed. For the remaining values of $t$, the distortion attains its extrema at zeros of the function of the form $$\left[-2\rho J_0(\rho/\lambda)+\lambda J_1(\rho/\lambda)\right]\cos(t/\lambda)\sin\theta\cos\theta\;.$$ If $[\dots]=0$ at some $\rho$, then the deformation at that location is independent of $\theta$. In all other cases, the extremal distorsions occur at $\theta_{ex}\in\{0,\pi/2,\pi,-\pi/2\}$, which demonstrates that  Halilsoy solution with $\alpha=0$ (i.e., the Einstein-Rosen standing wave) is linearly polarized, as expected.

\begin{figure}[t!]
        \includegraphics[width=0.945\linewidth,angle=0]{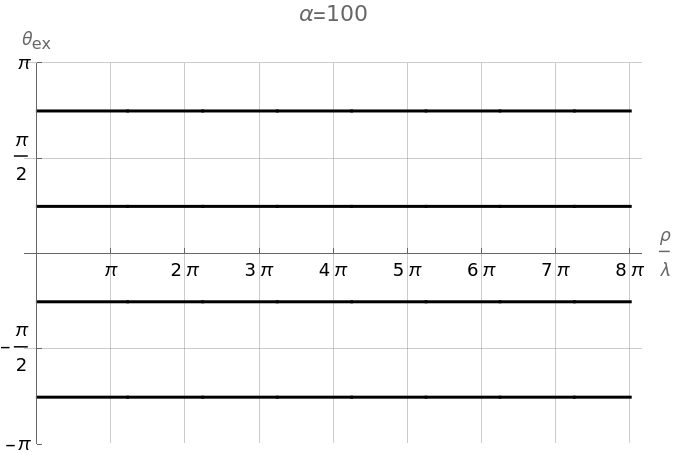}
	\caption{The angles of extremal distortion $\theta_{ex}$ for the Halilsoy standing gravitational wave for the large parameter $\alpha=100$. The cross polarization of the wave is clearly visible. The snapshot does not depend on $t$.} 
\label{fig4}
\end{figure}

For large $\alpha$ the term $$-\frac{1}{2}\epsilon J_1(\rho/\lambda)\sin(t/\lambda)\sin2\theta\sinh\alpha$$ in Eq.\ \eqref{deltaS} dominates the distortion away from its zeros. This term is extremal for $\cos2\theta=0$ so $\theta_{ex}$ can be approximated by $\theta_{ex}\in\{\pi/4,3\pi/4,-\pi/4,-3\pi/4\}$ which corresponds to the cross polarization. Therefore, the parameter $\alpha$ is related to the amplitude of a cross polarized component of the standing Halilsoy wave. This can be directly seen in Fig.\  \ref{fig4}.

\begin{figure}[t!]
        \includegraphics[width=0.945\linewidth,angle=0]{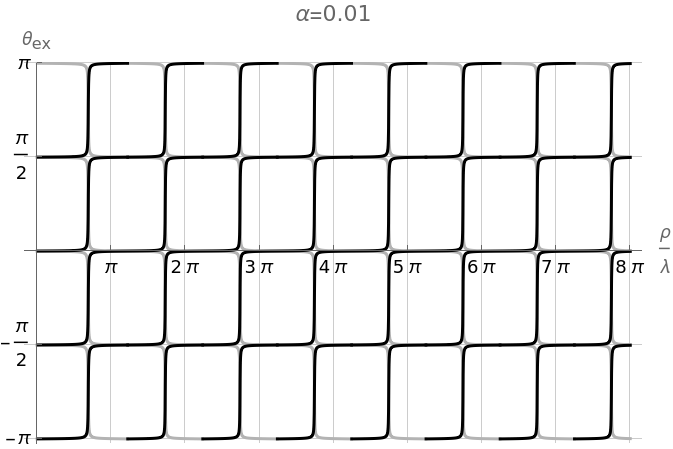}
	\caption{The angles of extremal distortion $\theta_{ex}$ for the Halilsoy standing gravitational wave for the small parameter $\alpha=0.01$. The snapshots are taken at $t/\lambda=7\pi/4$ (black curves) and $t/\lambda=5\pi/4$ (gray curves).}
\label{fig5}
\end{figure}

\begin{figure}[t!]
        \includegraphics[width=0.945\linewidth,angle=0]{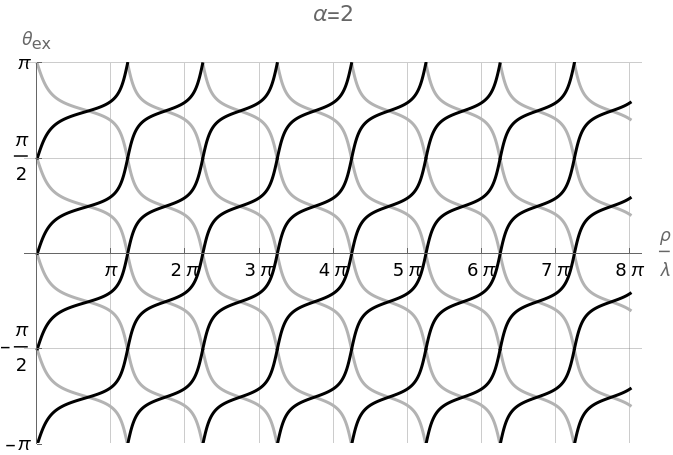}
	\caption{The angles of extremal distortion $\theta_{ex}$ for the Halilsoy standing gravitational wave for the parameter $\alpha=2$.  The snapshots are taken at $t/\lambda=7\pi/4$ (black curves) and $t/\lambda=5\pi/4$ (gray curves).}
\label{fig6}
\end{figure}

The intermediate values of the parameter $\alpha$ leads to more complicated behavior. This is a result of a nontrivial mixing of two polarization modes. Figures \ref{fig5}, \ref{fig6} show the angles of extremal distortions $\theta_{ex}$ for the Halilsoy standing wave for $\alpha=0.01$ and $\alpha=2$, respectively. The small nonzero parameter $\alpha$ induces a sharp transition, wherein the polarization deviates from the plus mode in a localized region. The growing $\alpha$ extends this transition region which, for large $\alpha$, covers almost the whole space and the cross polarization becomes dominant. 

\begin{figure}[t!]
        \includegraphics[width=0.945\linewidth,angle=0]{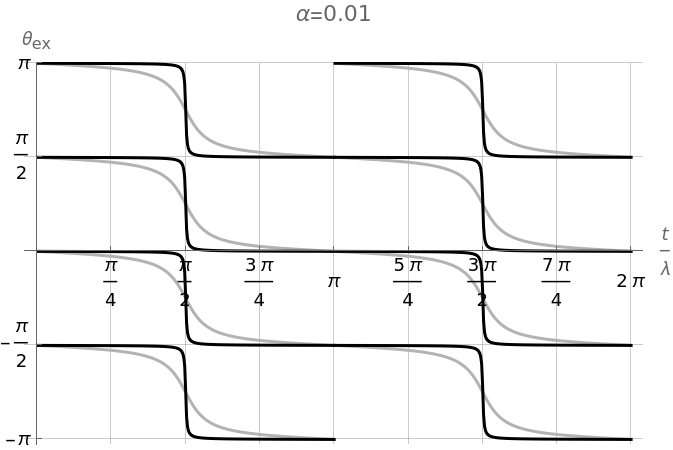}
	\caption{The angles of extremal distortion $\theta_{ex}$ for the Halilsoy standing gravitational wave for the small parameter $\alpha=0.01$.  The time evolution is presented at $\rho/\lambda=\pi/3$ (black curves) and $\rho/\lambda=2\pi/3$ (gray curves).}
\label{fig7}
\end{figure}

\begin{figure}[t!]
        \includegraphics[width=0.945\linewidth,angle=0]{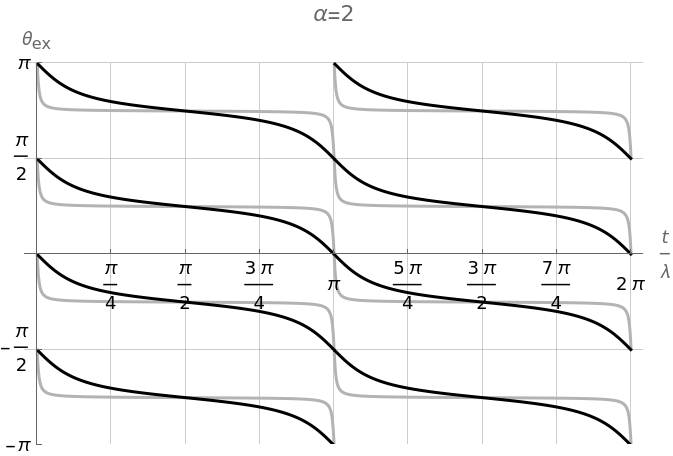}
	\caption{The angles of extremal distortion $\theta_{ex}$ for the Halilsoy standing gravitational wave for the parameter $\alpha=2$.  The time evolution is presented at $\rho/\lambda=\pi/3$ (black curves) and $\rho/\lambda=2\pi/3$ (gray curves).}
\label{fig8}
\end{figure}

The time evolution of the polarization follows similar pattern. We choose random timelike hypersurfaces \mbox{$\rho=\pi/3$}, \mbox{$\rho=2\pi/3$} and plot $\theta_{ex}$ as a function of $t/\lambda$ --- Figs.\ \ref{fig7}, \ref{fig8} for $\alpha=0.01$ and $\alpha=2$, respectively. 

In summary, the Halilsoy solution is plus polarized for $\alpha=0$ (the Einstein-Rosen standing wave). For large $\alpha$ the cross polarization dominates. For intermediate values of $\alpha$ the direction of extremal distortion of the test tube evolves in space and time in a way presented in  Figs.\ \ref{fig5}, \ref{fig6}, \ref{fig7}, \ref{fig8}. The complicated behavior of the polarization for the intermediate values of $\alpha$ can be explained by the following analysis in the asymptotic region $x\gg\lambda$ near a spatial infinity.

We consider superposition of two linearized plane waves. Both waves are linearly polarized with plus and cross polarization terms. They have the same amplitude, but they are moving in the opposite directions along the \mbox{$x$-axis}. In the TT gauge, the first wave is given by the formula
\begin{equation}\label{hTThat}
\begin{split}
        \hat{h}^{TT}_{yy}=&-\hat{h}^{TT}_{zz}=\epsilon\sin(\frac{t-x}{\lambda}-\pi/4)\;,\\
        \hat{h}^{TT}_{yz}=&\hat{h}^{TT}_{zy}=\epsilon\sin(\frac{t-x}{\lambda}-\pi/4)\sinh\alpha\;.\\
\end{split}
\end{equation}
The second wave moves in the opposite direction which is equivalent to the substitution $t\rightarrow -t$ in the equations above. In addition to that the cross term of the second wave is shifted in a phase by $\pi$. In the TT gauge, the second wave takes the form
\begin{equation}\label{hTThat2}
\begin{split}
        \hat{h}^{TT}_{yy}=&-\hat{h}^{TT}_{zz}=\epsilon\sin(\frac{-t-x}{\lambda}-\pi/4)\;,\\
        \hat{h}^{TT}_{yz}=&\hat{h}^{TT}_{zy}=-\epsilon\sin(\frac{-t-x}{\lambda}-\pi/4)\sinh\alpha\;.\\
\end{split}
\end{equation}

The superposition of these two waves forms a standing wave and leads to the distortion of the tube given by the formula
\begin{equation}\label{deltaS1}
	\begin{split}
		\Delta s(\phi=\pi/2)=&r\left[1+\epsilon\cos\frac{t}{\lambda}\sin(\frac{x}{\lambda}+\pi/4)\cos2\theta\right.\\
		&\left. +\epsilon\sin\frac{t}{\lambda}\cos(\frac{x}{\lambda}+\pi/4)\sin2\theta\sinh\alpha\right]\;.
	\end{split}
\end{equation}
We compare this formula to the one which corresponds to the Halilsoy solution---Eq.\ \eqref{deltaS} with $K=\sinh\alpha W$. Assuming $x\gg\lambda$, we find
\begin{equation}
\begin{split}
	\Delta s(\phi=\pi/2)\simeq& r\left[1+\epsilon\sqrt{\frac{\lambda}{2\pi x}}\cos\frac{t}{\lambda}\sin(\frac{x}{\lambda}+\pi/4)\cos2\theta\right.\\
	&\left. +\epsilon\sqrt{\frac{\lambda}{2\pi x}}\sin\frac{t}{\lambda}\cos(\frac{x}{\lambda}+\pi/4)\sin2\theta\sinh\alpha\right]\;.
\end{split}
\end{equation}
The formulas are identical up to the factor $\sqrt{\frac{\lambda}{2\pi x}}$ corresponding to the amplitude decay with distance from the center due to the spreading of energy over a cylindrical surface. 

Our analysis shows that the linearized Halilsoy standing wave far from the center corresponds to, using the terminology borrowed from electrodynamics, the ``polarization standing wave'' \cite{Fang2016}. It is a special case of this type of standing wave---the superposition of two orthogonally linearly polarized waves of equal amplitude propagating in opposite directions. These waves are not identical---the cross polarization term is shifted in a phase by $\pi$.

Polarization waves in electrodynamics have a constant energy density. Similarly, if we neglect the amplitude decay due to cylindrical symmetry, the C-energy of the linearized Halilsoy solution remains constant in the asymptotic region $x\gg\lambda$. In contrast, near the center the C-energy fluctuates, giving rise to an ``energy wave component''---analogous to that in electrodynamics \cite{Fang2016}---which vanishes with distance.

The polarization of the Chandrasekhar solution is more straightforward to analyze because factorization \eqref{fac} holds for all times.
Moreover, there is no free parameter to alter polarization. 
The factorization \eqref{fac} implies that $\theta_{ex}$ vary along the $x$-axis (assuming $\phi=\pi/2$), but not in time. Similarly to the Halilsoy solution, the linearized Chandrasekhar solution corresponds to the polarization standing wave, but it is a different type of this kind of a wave. This can be shown as follows.

Let us consider superposition of two linearized plane waves with a circular polarization of the same handedness. We assume that these waves are moving in the opposite directions along $x$-axis. We define these waves in the TT gauge by the following formulas. Nonzero components of the first wave are
\begin{equation}\label{hTTtilde}
\begin{split}
	\tilde{h}^{TT}_{yy}=&-\tilde{h}^{TT}_{zz}=\epsilon\sin(\frac{t-x}{\lambda}-\pi/4)\;,\\
	\tilde{h}^{TT}_{yz}=&\tilde{h}^{TT}_{zy}=\epsilon\sin(\frac{t-x}{\lambda}+\pi/4)\;.\\
\end{split}
\end{equation}
The second wave is identical, but it moves in the opposite direction which is equivalent to the substitution $t\rightarrow -t$. The superposition of these two waves forms polarization standing wave. The resulting distortion of the tube is given by the formula
\begin{equation}\label{deltaS2}
	\Delta s(\phi=\pi/2)=r\left[1+\epsilon\cos\frac{t}{\lambda}\sin(\frac{x}{\lambda}+2\theta+\pi/4)\right]\;.
\end{equation}
The equation above should be compared to the asymptotic form of the formula \eqref{deltaS} (assuming $x\gg\lambda$) for the Chandrasekhar solution (i.e., with $K=Q_1$)
\begin{equation}
	\Delta s(\phi=\pi/2)\simeq r\left[1+\epsilon\sqrt{\frac{\lambda}{2\pi x}}\cos\frac{t}{\lambda}\sin(\frac{x}{\lambda}+2\theta+\pi/4)\right]\;.
\end{equation}
Again, the formulas are identical up to the factor $\sqrt{\frac{\lambda}{2\pi x}}$ related to the amplitude decay in the cylindrical symmetry. Similarly to the Halilsoy solution, for the Chandasekhar solution the C-energy oscillates near the symmetry center but these oscillations vanish with distance.
Far from the center the linearized Chandrasekhar solution corresponds to the polarization standing wave which is a superposition of two circularly polarized waves of the same amplitude and handedness moving in the opposite directions.

\begin{figure}[t!]
        \includegraphics[width=0.945\linewidth,angle=0]{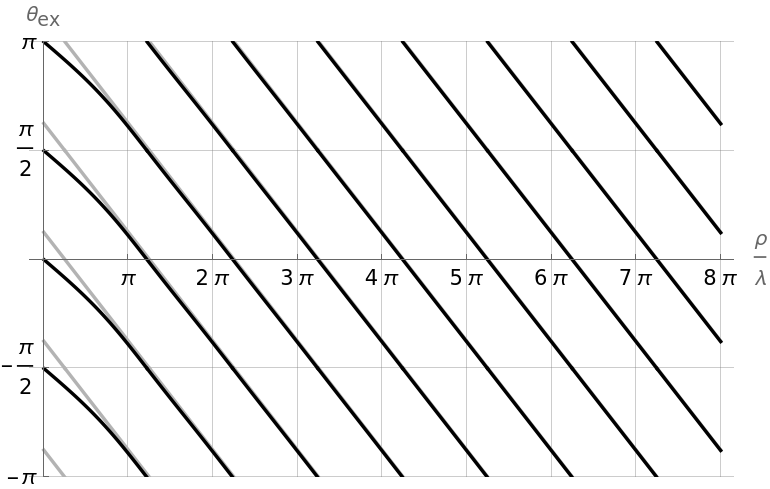}
	\caption{The angles of extremal distortion $\theta_{ex}$ for the Chandrasekhar standing gravitational wave (black curves). The snapshot is taken at $t/\lambda=7\pi/4$. Away from the center the direction of polarization varies linearly with the distance from the center. In the background $\theta_{ex}$ for linearized plane polarization wave is presented (gray curves). Both families of curves differ near the symmetry axis. They converge for larger $\rho$.}
\label{fig9}
\end{figure}

From a different point of view, one may say that the linearized Chandrasekhar standing wave is linearly polarized with the direction of polarization varying with the coordinate distance $\rho$ from the center. (In this picture, the linearized Halilsoy solution is circularly polarized except at isolated timelike hypersurfaces $J_1(\rho/\lambda)=0$ where the solution is linearly polarized and has only the plus mode.) The directions of deformation of the test tube are presented, for the Chandrasekhar solution, in \mbox{Fig.\ \ref{fig9}}. The gray curves correspond to the solution \eqref{hTTtilde}. As we move away from the center, the black curves corresponding to the Chandrasekhar solution converge toward the gray curves to eventually completely overlap them at larger $\rho$, in agreement with our analysis. Note, however, that the similarity between this figure and Fig.\ \ref{fig6} is misleading, since Fig.\ \ref{fig6} presents time-dependent snapshots.

\subsubsection{Nodes and antinodes}

The linearized solutions allows us to refine the definitions of  ``node'' and ``antinode'' in standing gravitational waves. According to a straightforward interpretation, nodes are the timelike hypersurfaces where the distortion of the test tube either vanishes or is minimal, whereas antinodes are the timelike hypersurfaces where the distortion is maximal. Ideally, the nodes would be determined by $t$ and $\theta$ independent zeros of the function $\Delta s(\hat n)|_{\phi=\pi/2}-r$ and the antinodes would be determined by $t$ and $\theta$ independent extremas of this function. The behavior the solutions studied in this paper is more complicated and these idealized assumptions, in general, are not satisfied even in the asymptotic region $\rho\gg\lambda$. The minima and extrema of $\Delta s$ depend on $\theta$, so the term ``maximal and minimal distortions'' needs clarification. 

The problem can be approached in several distinct ways. However, the most refined definition we developed for the solutions under investigation is to define the nodes and antinodes in terms of the circumference of a cross section of the test tube. This definition cannot be applied in the asymptotic region $\rho\gg\lambda$ because, in this limit, the circumference of a test tube become constant. Nevertheless, the analogy with polarization standing waves in electrodynamics gives us a clear understanding of the solutions far from the symmetry center, clarifying in which sense the linearized Halilsoy and Chandrasekhar solutions are standing gravitational waves there. 

In vacuum spacetimes, all curvature is encoded entirely in the Weyl tensor. Although the volumes of test spheres remain constant during evolution, neither the area nor circumference of their cross sections is conserved in general. Consequently, we propose ignoring deformations in other directions and defining nodes and antinodes by the circumference of the test tube's cross section in the direction orthogonal to the wave’s direction. Other definitions are possible, e.g.\ in terms of the area of the cross section of the tube, but the position of nodes and antinodes defined in this way may fluctuate and be time-independent only on average or in the asymptotic region. Thus, our definition best captures the sense in which the solutions under investigation can be regarded as standing waves.

For both linearized solutions studied in this paper the circumference of the cross section $C$ of the test tube is given by a remarkably simple formula
\begin{equation}
	C/r=2\pi+\epsilon\frac{\pi}{2}\frac{\lambda}{\rho}J_1(\rho/\lambda)\cos(t/\lambda)\;.
\end{equation}
The nodes and antinodes are then defined as the zeros and extrema of the function $C-2\pi r$ which corresponds to the zeros and extrema of the Bessel function $J_1(\rho/\lambda)$. For $\rho\gg\lambda$, we have $C/r\simeq 2\pi$ so the nodes and antinodes are not defined in this limit, as explained above.

Our definition of nodes and antinodes is slightly counter-intuitive. We illustrate this fact with the following example. 

\begin{figure}[t!]
        \includegraphics[width=0.945\linewidth,angle=0]{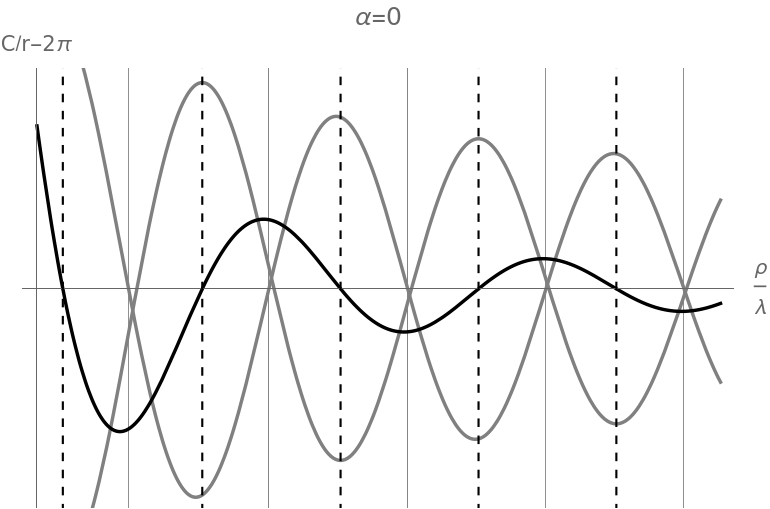}
	\caption{The spatial component of the distortion of the circumference of the test tube $C/r-2\pi$ (black) plotted against spatial components of distortions $\Delta s(\theta_{ex})/r-1$ (gray). The vertical lines correspond to zeros (dashed lines) and extrema (solid lines) of the Bessel function $J_1(\rho/\lambda)$. They define nodes (dashed lines) and antinodes (dotted lines) of standing waves for the linearized Halilsoy and Chandrasekhar solutions. The figure is plotted for the Halilsoy solution with $\alpha=0$, but it illustrates the general behavior of both solutions. The deformation is exaggerated for clarity ($\epsilon=1/2$).}
\label{fig10}
\end{figure}

Let us consider the Halilsoy solution with $\alpha=0$ (i.e., the Einstein-Rosen standing wave). It follows from the analysis presented in the previous section that this solution is linearly polarized and has only the plus mode, hence 
$$\theta_{ex}\in\{-\pi/2,0,\pi/2,\pi\}\;.$$ The four $\theta_{ex}$'s give rise to two distinct values of $\Delta s$, namely (for $\phi=\pi/2$)
\begin{equation}
\begin{split}
	\Delta s(\theta\in\{0,\pi\})=&1+\frac{\epsilon}{2}J_0(\rho/\lambda)\cos(t/\lambda)\;,\\
	\Delta s(\theta=\pm\pi/2)=&1+\frac{\epsilon}{2}\left(-J_0(\rho/\lambda)+\frac{\lambda}{\rho}J_1(\rho/\lambda)\right)\cos(t/\lambda)\;.
\end{split}
\end{equation}
We plot the spatial amplitudes of the distortions [neglecting the factor $\cos(t/\lambda)$] in Fig.\ \ref{fig10}. The spatial part of the distortion of the circumference of the tube vanishes roughly at places where the spatial components of the tube distortion have their extrema. Therefore, in our definition, the nodes correspond to places at which the test tube, at first glance, seems to attain its maximal distortion---a fact that is clearly counterintuitive. This happens because the distortions (e.g.\ at $\theta\in\{0,\pi\}$ and at $\theta=\pm\pi/2$) have different signs so their contribution to the distortion of the circumference of the tube cancels out. Similarly, the circumferential distortion is maximal at places where the amplitude of the distortions is small, but where they have the same sign. These locations are identified as antinodes in our definition. 

Our definitions of nodes and antinodes apply to the linearized solution, and it is not immediately clear how they extend to fully nonlinear solutions (the examples studied in this paper or other spacetimes). Nevertheless, we note that in the case of the full nonlinear Halilsoy solution, these definitions are related to C-energy. The hypersurfaces $J_1(\rho/\lambda)=0$, which we identify as nodes, are distinguished by the fact that C-energy remains strictly constant in time along these hypersurfaces. Moreover, the metric function $\omega$ vanishes there. In the linear regime, this results in linear polarization of the Halilsoy solution at the nodes, which serve, for this type of polarization wave, as boundaries between regions of circular polarization.

\section{Summary}

We discovered that the Halilsoy \eqref{metricH} solution and the Chandrasekhar solution \eqref{metricCh} are, in the linear approximation, gravitational analogs of of two types of electromagnetic polarization standing waves \cite{Fang2016}. This analogy holds strictly far from the symmetry center, where the fluctuations in C-energy vanish. Furthermore, we have clarified in which sense the solutions under investigation are standing gravitational waves near the symmetry center---the circumference of the test tube's cross section exhibits characteristic for standing waves behavior, with fixed positions of nodes and antinodes. 

In light of our results, the less restrictive definition of standing gravitational waves provided by Stephani \cite{Stephani:2003}---according to which C-energy must be constant only on average in time---appears to be sufficient. Although the Halilsoy solution does not satisfy the more restrictive Chandrasekhar definition \cite{chandra}, it still exhibits typical standing waves behavior in the weak-field approximation. 

It would be interesting to examine the high frequency limit of these solutions. According to the hypothesis presented in the article \cite{Szybka:2019}, the effective energy-momentum tensor should have a particular form.

The interpretation we have discovered for the linearized Halilsoy and Chandrasekhar solutions is consistent with the behavior of C-energy. Therefore, we expect it to carry over to the full nonlinear solutions.

\section*{Acknowledgments}

Some calculations were performed using Wolfram {\it Mathematica} and the xAct package \cite{xAct2021}. We gratefully acknowledge the contributions of OpenAI's GPT and Copilot for their assistance in resolving linguistic nuances, formatting LaTeX formulas, and generating figures in {\it Mathematica}.

\section*{Data availability}

No data were created or analyzed in this study.

\bibliographystyle{apsrev4-2}
\setcitestyle{authortitle}
\bibliography{report}

\end{document}